\documentclass[11pt,draftcls,onecolumn]{IEEEtranwc}

\usepackage{color}
\usepackage{amsmath,amssymb,amscd}
\usepackage{subfigure, epsfig}
\newtheorem{lemma}{Lemma}
\newtheorem{theorem}{Theorem}
\newcommand{\mb}{\mathbf}
\newcommand{\ul}{\underline}
\newcommand{\ol}{\overline}
\newcommand{\bs}{\boldsymbol}
\newcommand{\mc}{\mathcal}

\newcommand{\ds}{\displaystyle}

\ifCLASSINFOpdf
\else
\fi

\begin{document}

\title{Power Allocation Games  for \\ MIMO Multiple Access Channels with Coordination}

\author{
Elena-Veronica~Belmega,~\IEEEmembership{Student~Member,~IEEE,}
Samson~Lasaulce,~\IEEEmembership{Member,~IEEE,}
        and~Merouane~Debbah,~\IEEEmembership{Senior~Member,~IEEE}

\thanks{The material in this paper was presented in part at the 6th
IEEE/ACM Intl. Symposium on Modeling and Optimization in Mobile, Ad
Hoc, and Wireless Networks and Workshops (WiOpt), Berlin, Germany, 4
April 2008 \cite{belmega-wnc3-2008}.}
\thanks{E.~V. Belmega and S. Lasaulce are with LSS (joint lab of CNRS, Sup\'{e}lec, Paris 11), Sup\'{e}lec,
Plateau du Moulon, 91192 Gif-sur-Yvette, France,
\{belmega,lasaulce\}@lss.supelec.fr; M. Debbah is with the ``Chaire
Alcatel-Lucent'' at Sup\'{e}lec, merouane.debbah@supelec.fr.}}

\maketitle

\begin{abstract}
A game theoretic approach is used to derive the optimal
decentralized power allocation (PA) in fast fading multiple access
channels where the transmitters and receiver are equipped with
multiple antennas. The players (the mobile terminals) are free to
choose their PA in order to maximize their individual transmission
rates (in particular they can ignore some specified centralized
policies). A simple coordination mechanism between users is
introduced. The nature and influence of this mechanism is studied in
detail. The coordination signal indicates to the users the order in
which the receiver applies successive interference cancellation and
the frequency at which this order is used. Two different games are
investigated: the users can either adapt their temporal PA to their
decoding rank at the receiver or optimize their spatial PA between
their transmit antennas. For both games a thorough analysis of the
existence, uniqueness and sum-rate efficiency of the network Nash
equilibrium is conducted. Analytical and simulation results are
provided to assess the gap between the decentralized network
performance and its equivalent virtual multiple input multiple
output system, which is shown to be zero in some cases and
relatively small in general.
\end{abstract}

\begin{IEEEkeywords}
Game theory, large systems, MAC, MIMO, Nash equilibrium, power
allocation games, random matrix theory.
\end{IEEEkeywords}

\section{Introduction} \label{sec:intro}
\IEEEPARstart{W}{e} consider a special case of decentralized or
distributed wireless networks, the decentralized multiple access
channel (MAC). In this context,  the MAC consists of a network of
several mobile stations (MS) and one base station (BS). In the
present work, the network is said to be decentralized in the sense
that each user can choose freely his power allocation (PA) policy in
order to selfishly maximize a certain individual performance
criterion. This means that, even if the the BS broadcasts some
specified policies, every (possibly cognitive) user is free to
ignore the policy intended for him if the latter does not maximize
his performance criterion.

The problem of decentralized PA in wireless networks is not new and
has been properly formalized for the first time in
\cite{grandhi-allerton-1992, grandhi-comm-1994}. Interestingly, this
problem can be formulated quite naturally as a non-cooperative game
with  different performance criteria (utilities) such as the
carrier-to-interference ratio \cite{ji-wn-1998}, aggregate
throughput \cite{oh-infocom-2000} or energy efficiency
\cite{goodman-pcomm-2000}, \cite{meshkati-jsac-2006}. In this paper,
we assume that the users want to maximize information-theoretic
utilities and more precisely their Shannon transmission rates. 
Many reasons why this kind of utilities is often considered are
provided in the literature related to the problem under
investigation (some references are provided further). Here we will
just mention three of them. First, Shannon transmission rates allow
one to characterize the performance limits of a communication system
and study the behavior of (selfish) users in a network where good
coding schemes are implemented. As there is a direct relationship
between the achievable transmission rate of a user and his
signal-to-interference plus noise ratio (SINR), they also allow one
to optimize performance metrics like the SINR or related quantities
of the same type (e.g., the carrier-to-interference ratio). From the
mathematical point of view, Shannon rates have many desirable
properties (e.g., concavity properties), which allows one to conduct
deep performance analyses. Therefore they provide useful insights
and concepts that are exploitable for a practical design of
decentralized networks. Indeed, the point of view adopted here is
close to the one proposed by the authors of \cite{yu-jsac-2002} for
DSL (digital subscriber lines) systems, which are modeled as a
parallel interference channel; \cite{lai-it-2008} for the single
input single output (SISO) and single input multiple output (SIMO)
fast fading MACs with global CSIR and global CSIT (Channel State
Information at the Receiver/Transmitters);
\cite{lasaulce-gamecomm-2007} for MIMO (Multiple Input Multiple
Output) MACs with global CSIR, channel distribution information at
the transmitters (global CDIT) and single-user decoding (SUD) at the
receivers; \cite{arslan-wc-2007, scutari-jsac-2008} for Gaussian
MIMO interference channels with global CSIR and local CSIT and, by
definition of the conventional interference channel
\cite{carleial-it-1978}, SUD at the receivers. Note that reference
\cite{palomar-it-2003} where the authors considered Gaussian MIMO
MACs with neither CSIT nor CDIT differs from our approach and that
of \cite{yu-jsac-2002, lai-it-2008, lasaulce-gamecomm-2007,
arslan-wc-2007, scutari-jsac-2008} because in \cite{palomar-it-2003}
the MIMO MAC is seen as a two-player zero-sum game where the first
player is the group of transmitters and the second player is the set
of MIMO sub-channels. In the list of the aforementioned references,
\cite{lai-it-2008} seems to be the closest work to ours. However,
our approach differs from \cite{lai-it-2008} on several technical
key points. First of all, not only the BS but also the MSs can be
equipped with multiple antennas. This is an important technical
difference since the power control problem of \cite{lai-it-2008}
becomes a PA problem for which the precoding matrix of each user
has to be determined. 
Also the issues regarding the existence and uniqueness of the
network equilibrium
 are more complicated to be dealt with, as it will be seen.
 Specifically, random matrix theory will be exploited to
determine the optimum eigenvalues of the precoding matrices. In
\cite{lai-it-2008}, several assumptions made, especially the one involving the
knowledge of all the instantaneous channels at each MS can be argued
in some contexts. One of our objectives is to decrease the amount of
signaling needed from the BS. This is why we assume that the BS can
only send to the users sufficient training signals for them to
know the statistics of the different channels and a simple and
common coordination signal. The underlying coordination mechanism is
simple because it consists in periodically sending the realization
of a $K!$-state random signal, where $K$ is the number of active
users. As it will be seen in detail, such a mechanism is mandatory
because, in contrast with \cite{lasaulce-gamecomm-2007}, we assume
here successive interference cancellation (SIC) at the BS. Thus each
user needs to know his decoding rank in order to adapt his PA policy to maximize the
transmission rate. The coordination signal precisely indicates to
all the users the decoding order employed by the receiver. Therefore the
proposed formulation can be seen from two different standpoints. If
the distribution of the coordination signal is fixed, then the
addressed problem can be regarded as a non-cooperative game where the BS
is imposed to follow the realizations of the random coordination
signal. In this case the respective signal can be generated by any device (and not necessarily by
the BS), in order to select the decoding order. On the other hand,
if the distribution of the coordination signal can be optimized, the
problem can be addressed as a Stackelberg game. Here the BS is the game
leader and chooses his best mixed strategy (namely a distribution
over the possible decoding orders) in order to maximize a certain
utility, which will be chosen to be the network uplink sum-rate.

In the described framework, one of our objectives is to know how
well a non-cooperative but weakly coordinated system performs in
terms of overall sum-rate w.r.t. its centralized counterpart (by
``centralized'' we mean that the users are imposed to follow the BS
PA policies) when SIC is used at the BS. In this setting, several
interesting questions arise. When the users' utility functions are
chosen to be their individual transmission rates, is there a Nash
equilibrium (NE) in the corresponding game and is it unique? What is
the optimum way for a selfish user to allocate (spatially or temporally) his transmit
power? How to choose the coordination signal that
maximizes the network sum-rate? What is the performance loss of the
decentralized network w.r.t. the equivalent virtual MIMO network?

This paper is structured as follows. After presenting the system
model (Sec. \ref{sec:system-model}), we study in detail two PA
games. In the first case (Sec. \ref{sec:temporal-PA-game}), each MS
is imposed to share his power uniformly between his transmit
antennas but can freely allocate his power over time. In the second
case (Sec. \ref{sec:spatial-PA-game}), we assume that the temporal
PA is uniform and thus our objective is to derive the best spatial
PA scheme. For each of these frameworks the existence, uniqueness,
determination and sum-rate efficiency of the NE is investigated.
Numerical results are provided in Sec. \ref{sec:simulation-results}
to illustrate our theoretical analysis and in particular to better
assess the sum-rate efficiency of the different games considered. We
conclude the paper by several remarks and possible extensions of our
work in Sec. \ref{sec:conclusions}.

\section{System Model} \label{sec:system-model}

Throughout the paper $\ul{v}$, $\mb{M}$, $(.)^T$ and $(.)^H$ will stand
for vector, matrix, transpose and transpose conjugate, respectively.
For simplicity and without loss of generality, we will assume a MAC
with $K=2$ users. Note that the type of multiple access technique
assumed corresponds to the one considered in the standard definition
of the Gaussian MAC by \cite{wyner-it-1974},\cite{cover-book-1975}:
all transmitters send at once and at different rates over the entire
bandwidth. In this (information theoretic) context, very long
codewords can be used and the receiver is not limited in terms of
complexity. Thus the codewords of the different transmitters can be
decoded jointly using a maximum likelihood decoding procedure (see
\cite{cover-book-1975} for more details). Interestingly, the
transmission rates of the capacity region corresponding to the
coding-decoding procedure just mentioned, can also be achieved, as
discussed in \cite{cover-book-1975}, by using perfect SIC at the
receiver. In this paper we also adopt this decoding scheme, which
means that not only the different channel matrices are perfectly
known to the receiver but also that the codewords of all the users
are decoded reliably. The case of imperfect CSIR and error
propagation in the SIC procedure is thus seen as a useful
extension of this paper. Since we assume SIC at the BS and that the
users want to maximize their individual transmission rates, it is
necessary for them to know the decoding order used by the BS. This
is why we assume the existence of a source broadcasting a discrete
coordination signal to all the terminals in presence. If this source
is the BS itself, this induces a certain cost in terms of downlink
signaling but the distribution of the coordination signal can then
be optimized. On the other hand, if the coordination signal comes
from an external source, e.g., an FM transmitter, the MSs can
acquire their coordination signal for free in terms of downlink
signaling. However this generally involves a certain sub-optimality in
terms of uplink rate. Analyzing this kind of tradeoffs is precisely
one of the goals of this paper. In both cases, the coordination
signal will be represented by a Bernouilli random variable denoted
with $S \in \mc{S}$. Since we study the $2-$user MAC, $\mc{S} = \{1,2\}$ is a binary alphabet
and $S$ is distributed as
$\mathrm{Pr}[S=1] = p$, $\mathrm{Pr}[S=2] = 1-p \triangleq \ol{p}$.
Without loss of generality we assume that when the realization of
$S$ is $1$, user 1 is decoded in the second place and therefore sees
no multiple access interference; in a real wireless system the
frequency at which the realizations would be drawn is roughly
proportional to the reciprocal of the channel coherence time
($T_{\mathrm{coh}}$). Note that the proposed coordination mechanism
is suboptimal in the sense that the coordination signal does not
depend on the realizations of the channel matrices. We will see that
the corresponding performance loss is in fact very small.

We will further consider that each MS is equipped with $n_t$
antennas whereas the BS has $n_r$ antennas. In our analysis, the
flat fading channel matrices of the different links vary from symbol
vector to symbol vector. We assume that the receiver knows all the
channel matrices whereas the transmitters have only access to the
statistics of the different channels. At this point, the authors
would like to re-emphasize their point of view:
\begin{itemize}
    \item On the one hand, we think that in some contexts our approach can be
interesting in terms of signaling cost. We have seen that $S$ lies
in a $K-$element alphabet and the realizations are drawn
approximatively at $\frac{1}{T_{\mathrm{coh}}}\mathrm{[Hz]}$,
therefore the coordination mechanism requires at most
$\frac{\log_2(K!)}{T_{\mathrm{coh}}}$ bps from the BS and $0$ bps if
it is built from an external source. Another source of signaling
cost is the acquisition of the knowledge of the statistics of the
uplink channels at the MSs. For example, in the context of coherent
communications where the BS regularly sends some data to the MSs and
channel reciprocity assumption is valid (e.g., in time division
duplex systems) the corresponding cost can be reasonable. In
general, this cost will have to be compared to the cost of the
centralized system where the BS has to send accurate enough
quantized versions of the (possibly large) precoding matrices at a
certain frequency.
    \item On the other hand, even if our approach is not interesting in terms
of signaling, it can be very useful in contexts where terminals are
autonomous and may have some selfish reasons to deviate from the
centralized policies. In such scenarios, the concept of network
equilibrium is of high importance.
\end{itemize}

The equivalent baseband signal received by the BS can be written as:
\begin{equation}\label{eq:system-model-mimo}
\ul{y}^{(s)}(\tau)  =\sum_{k=1}^K \bs{H}_k(\tau)
\ul{x}_k^{(s)}(\tau) + \ul{z}^{(s)}(\tau),
\end{equation}
where $\ul{x}_k^{(s)}(\tau)$ is the $n_t$-dimensional column vector
of symbols transmitted by user $k$ at time $\tau$ for the
realization $s \in \mc{S}$ of the coordination signal,
$\mb{H}_k(\tau) \in \mathbb{C}^{n_r \times n_t} $ is the channel
matrix (stationary and ergodic process) of user $k$ and
$\ul{z}^{(s)}(\tau)$ is an $n_r$-dimensional complex white Gaussian
noise distributed as $\mathcal{N}(\ul{0}, \sigma^2
\mathbf{I}_{n_r})$; for sake of clarity we will omit the time index
$\tau$ from our notations. As \cite{telatar-ett-1999} we assume
that, for each $s \in \mathcal{S}$, the data streams of user
$k$ are multiplexed in the eigen-directions of the matrix
$\mb{Q}_k^{(s)} = \mathbb{E} \left[\ul{x}_k^{(s)} \ul{x}_k^{(s),H}
\right] \triangleq \mb{V}_k^{(s)} \mb{P}_k^{(s)}\mb{V}_k^{(s),H}$.
Finding the optimal eigen-values $\mb{P}_k^{(s)}$ and coordinate
systems $\mb{V}_k^{(s)}$ that maximize the transmission rate of user $k$ is one of the main issues we will solve in the next two
sections. In order to take into account the antenna correlation
effects at the transmitters and receiver, we assume the different
channel matrices to be structured according to the Kronecker
propagation model \cite{shiu-com-2000} with common receive
correlation \cite{taricco-globecom-2007}:
\begin{equation}
\label{eq:kronecker-model} \forall k \in \{1,...,K\}, \ \mb{H}_k =
\mb{R}^{\frac{1}{2}} \bs{\Theta}_k \mb{T}_k^{\frac{1}{2}}
\end{equation}
where $\mb{R}$ is the receive antenna correlation matrix, $\mb{T}_k$
is the transmit antenna correlation matrix for user $k$ and
$\bs{\Theta}_k$ is an $n_r \times n_t$ matrix whose entries are
zero-mean independent and identically distributed complex Gaussian
random variables with variance $\frac{1}{n_t}$. The motivation for
assuming a channel model with common receive correlation is twofold.
First, there exist some situations where this MIMO MAC model is
realistic, the most simple situation being the case of no receive
correlation i.e., $\mb{R} = \mb{I}$ (see e.g.,
\cite{soysal-tcsubmitted-2007}). Although it is not explicitly
stated in \cite{taricco-globecom-2007} the second feature of this
model is that the overall channel matrix $\mb{H} = \left[\mb{H}_1
... \mb{H}_K \right]$ can also be factorized as a Kronecker model,
which will allow us to re-exploit existing results from the random
matrix theory literature. Therefore the case where the overall
channel matrix is not separable can be seen as a possible extension
of this paper that can be dealt with by using the results in
\cite{girko-book-2001}.

In this paper we study in detail two special but useful cases of
decentralized PA problems. In the first case (Game 1), we assume
(for instance because of practical technical/complexity constraints)
that each user is imposed to share his power uniformly between his
transmit antennas but can freely allocate his power over time; this
problem will be referred to as temporal PA game (Sec.
\ref{sec:temporal-PA-game}). In the second case (Game 2), for every
realization of the coordination signal, each user is assumed to
transmit with the same total power (denoted by $P_k$) but can freely
share it between his antennas; this problem will be referred to as
spatial PA game (Sec. \ref{sec:spatial-PA-game}). For both games the
\emph{strategy} of user $k \in \{1,2\}$ consists in choosing the
distribution of $\ul{x}_k^{(s)}$, for each $s\in \mc{S}$ in order to
maximize his \emph{utility} function which is given by:
\begin{equation}
\label{eq:utility-mimo}
\begin{array}{ccl}
u_k(\mb{Q}_1^{(1)}, \mb{Q}_1^{(2)}, \mb{Q}_2^{(1)}, \mb{Q}_2^{(2)})
&=& \ds{\sum_{s=1}^{2} \mathrm{Pr}[S=s] R_k^{(s)} (\mb{Q}_1^{(s)},
\mb{Q}_2^{(s)})}
\end{array}
\end{equation}
where
\begin{equation}
\label{eq:mimo-st-rates} R_k^{(s)}(\mb{Q}_1^{(s)},\mb{Q}_2^{(s)}) =
\left|
\begin{array}{ll}
 \mathbb{E} \log | \mb{I} + \eta \mb{H}_k \mb{Q}_k^{(s)}
\mb{H}_k^H | & \mathrm{if} \
k = s\\
\mathbb{E} \log | \mb{I} + \eta \sum_{k=1}^{2} \mb{H}_k
\mb{Q}_k^{(s)} \mb{H}_k^H|- \mathbb{E} \log |\mb{I} + \eta
\mb{H}_{-k} \mb{Q}_{-k}^{(s)} \mb{H}_{-k}^H| & \mathrm{if} \ k \neq
s
\end{array}
\right.
\end{equation}
with $\eta \triangleq \frac{1}{\sigma^2}$ and the usual notation for
$-k$, which stands for the other user than $k$. Note that we
implicitly assume Gaussian codebooks for the two users since this
choice is optimum in terms of their individual Shannon transmission
rates (see e.g., \cite{tse-book-2005}). This is why the strategy of
a user boils down to choosing the best pair of covariance matrices
$(\mb{Q}_k^{(1)}, \mb{Q}_k^{(2)})$. The corresponding maximization
is performed under the following transmit power constraint for each
MS: $\mathrm{Tr} \left(  \sum_{s=1}^{2} \mathrm{Pr}[S=s]
\mb{Q}_k^{(s)} \right) \leq n_t P_k$. The main difference between
Games 1 and 2 relies precisely on how this general power constraint
is specialized. In Game 1, the precoding matrices are imposed to
have the following structure: $\forall k \in \{1,2\}, \forall s \in
\{1,2 \}, \mb{Q}_k^{(s)} = \alpha_k^{(s)} P_k \mb{I}_{n_t}$, which
amounts to rewriting the total power constraint as follows
\begin{equation}
\label{eq:constraint-temporal-pa} \sum_{s=1}^{2} \mathrm{Pr}[S=s]
\alpha_k^{(s)} \leq 1.
\end{equation}
On the other hand, in Game 2, the power constraint expresses as
\begin{equation}
\label{eq:constraint-spatial-pa} \forall k \in \{1,2\}, \forall s
\in \{1,2 \}, \mathrm{Tr}(\mb{Q}_k^{(s)}) \leq n_t P_k.
\end{equation}
In both game frameworks, an important issue for a wireless network
designer/owner is to know whether by leaving the users decide their PA by
themselves, the network is going to operate at a given and
predictable state. This precisely corresponds to the notion of a
network equilibrium, a state from which no user has interest to
deviate. The main issue is to know if there exists an equilibrium
point, whether it is unique, how to determine the corresponding
strategies and characterize the efficiency of this equilibrium in
terms of network sum-rate.

\section{Temporal power allocation game}
\label{sec:temporal-PA-game}
%
%
As mentioned above, in the temporal power allocation (TPA) game, the
strategy of user $k \in \{1,2\}$ merely consists in choosing the
best pair $(\alpha_k^{(1)}, \alpha_k^{(2)})$. Since each
transmission rate is a concave and non-decreasing function of the
$\alpha_k^{(s)}$'s, each user will saturate the power constraint (5)
i.e., $\sum_{s=1}^{2} \mathrm{Pr}[S=s] \alpha_k^{(s)} = 1$, which
leads to optimizing a single parameter $\alpha_k^{(1)}$ or
$\alpha_k^{(2)}$. From now on, for sake of clarity we will use the
notations $\alpha_1^{(1)} = \alpha_1$, $\alpha_2^{(2)} = \alpha_2$.
Indeed, it is easy to verify that the power constraints are
characterized completely, for the first user by
$\alpha_1^{(2)}=\frac{1-p\alpha_1}{1-p}$ with $\alpha_1\in
\mc{A}_1^{\mathrm{TPA}} \triangleq \left[0,\frac{1}{p} \right]$, and
for the second user by $\alpha_2^{(1)}=\frac{1-(1-p)\alpha_2}{p}$
with $\alpha_2 \in \mc{A}_2^{\mathrm{TPA}} \triangleq \left[0,
\frac{1}{1-p} \right]$. Thus the strategy of user $k \in
\{1,2\}$ consists in choosing the best fraction $\alpha_k$ from the action 
set $\mc{A}_k^{\mathrm{TPA}}$. Our main goal is to
investigate if there exists an NE and determine the corresponding
profile of strategies $(\alpha_1^{\mathrm{NE}},
\alpha_2^{\mathrm{NE}})$. It turns out that the issues of the
existence and uniqueness of an NE can be properly dealt with by
applying Theorems 1 and 2 of \cite{rosen-eco-1965} in our context.
For making this paper sufficiently self-contained, we review here
these two theorems (Theorem 2 is given for the $2-$user case for
simplicity and because it is sufficient under our assumptions).
\begin{theorem}\cite{rosen-eco-1965} \emph{Let $\mc{G} = (\mc{K}, \{\mc{A}_k\}_{k\in\mc{K}},\{u_k\}_{k\in\mc{K}}) $ be a game
where $\mc{K} = \{1,...,K\}$ is the set of players,
$\mc{A}_1,...,\mc{A}_K$ the corresponding sets of strategies and
$u_1,...,u_k$ the utilities of the different players. If the
following three conditions are satisfied: (i) each $u_k$ is
continuous in the vector of strategies $(a_1,...,a_K) \in
\prod_{k=1}^{K} \mc{A}_k$; (ii) each $u_k$ is concave in $a_k \in
\mc{A}_k$; (iii) $\mc{A}_1, ..., \mc{A}_K$ are compact and convex
sets; then $\mc{G}$ has at least one NE.} \label{theo-1-rosen}
\end{theorem}

\begin{theorem}\cite{rosen-eco-1965} \emph{Consider the $K$-player concave game of Theorem
\ref{theo-1-rosen} with $K=2$. If the following (diagonally strict
concavity) condition is met: for all $(a_1',a_1'') \in \mc{A}_1^2$
and $(a_2',a_2'') \in \mc{A}_2^2$ such that $(a_1', a_2') \neq
(a_1'', a_2''), (a_1''-a_1') \left[ \frac{\partial u_1}{\partial
a_1} (a_1',a_2') - \frac{\partial u_1}{\partial a_1}
(a_1'',a_2'')\right]+(a_2''-a_2') \left[ \frac{\partial
u_2}{\partial a_2} (a_1',a_2') - \frac{\partial u_2}{\partial a_2}
(a_1'',a_2'')\right]  >0$; then the uniqueness of the NE is
insured.} \label{theo-2-rosen}
\end{theorem}
At this point we can state the first main result of this paper,
which is provided in the following theorem. For sake of clarity we will
also use the notations: $p_k \triangleq p$ if $k=1$ or $p_k
\triangleq \ol{p}$ if $k=2$.

\begin{theorem}[Existence and uniqueness of an NE in Game 1]
\emph{the temporal PA game described by: the set of players $\mc{K}
= \{1,2\}$; the sets of actions $ \mc{A}_k^{\mathrm{TPA}} =\left[0,
\frac{1}{p_k} \right]$ and utilities $u_k(\alpha_k,\alpha_{-k}) = p
R_k^{(1)}(\alpha_k,\alpha_{-k})+\ol{p}R_k^{(2)}(\alpha_{k},\alpha_{-k})$,
where the rates $R_k^{(s)}$ follow from Eq. (\ref{eq:mimo-st-rates})
has a unique NE.}
\end{theorem}
\begin{proof}

\emph{Existence of an NE.} It is guaranteed by the geometrical and
topological properties of the utility functions and the strategy
sets of the users (over which the maximization is performed). Indeed,
we can apply \cite{rosen-eco-1965} in our matrix case. Without loss
of generality, let us consider user 1. The utility of user 1
comprises two terms corresponding to the two coordination signal
realizations: $u_1(\alpha_1, \alpha_2) = p R_1^{(1)}(\cdot,\cdot) +
\ol{p} R_1^{(2)}(\cdot,\cdot)$. Using the fact that $
\frac{d^2}{dt^2}\log |\mb{X}+ t\mb{Y}\mb{Y}^H| = -
\mathrm{Tr}\left[\mb{Y}^H(\mb{X}+
t\mb{Y}\mb{Y}^H)^{-1}\mb{Y}\mb{Y}^H(\mb{X}+
t\mb{Y}\mb{Y}^H)^{-1}\mb{Y}\right]$ it is easy to verify that $
\frac{\partial^2 R_1^{(1)}}{\partial \alpha_1^2} (\alpha_1,\alpha_2)
= - \mathbb{E} \mathrm{Tr} [\mb{B} \mb{B}^H] <0$ and
$\frac{\partial^2 R_1^{(2)}}{\partial \alpha_1^2}
(\alpha_1,\alpha_2) = - \mathbb{E} \mathrm{Tr} [\mb{C} \mb{C}^H] <0$
where $\mb{B} = \rho_1 \mb{H}_1^H (\mb{I}+\rho_1 \alpha_1 \mb{H}_1
\mb{H}_1^H )^{-1}\mb{H}_1$, $\mb{C}=\frac{p}{\ol{p}}\rho_1
\mb{H}_1^H \left(\mb{I}+\rho_1 \frac{1-p\alpha_1}{\ol{p}}
 \mb{H}_1 \mb{H}_1^H  + \rho_2 \alpha_2 \mb{H}_2
 \mb{H}_2^H\right)^{-1}\mb{H}_1$ and $\rho_1=\eta P_1$, $\rho_2=\eta P_2$ correspond to the
signal-to-noise ratios of the users. Thus for every user $k$, the utility $u_k$ is strictly concave
w.r.t. to $\alpha_k$. Also it is continuous in $(\alpha_1,
\alpha_2)$ over the convex and compact strategy sets
$\mc{A}_k^{\mathrm{TPA}}$. Therefore the existence of at least one
NE is guaranteed. Interestingly, we observe that for a fixed game
rule, which is the value of the parameter $p$, there will always be
an equilibrium. The users adapt their strategies to the rule of the
game in order to optimize their individual transmission rates.

\emph{Uniqueness of the NE.} We always apply \cite{rosen-eco-1965}
in our matrix case (see Appendix \ref{appendix_1}) and prove that
the diagonally strict concavity condition is actually met. The key
of the proof is the following Lemma which is proven in Appendix
\ref{appendix_2}.
\begin{lemma}\label{lemma:diago}
\emph{Let $\mb{A}'$, $\mb{A}''$, $ \mb{B}'$ and $\mb{B}''$ be
Hermitian and non-negative matrices such that either $\mb{A}' \neq
\mb{A}''$ or $\mb{B}' \neq \mb{B}''$. Assume that the classical
matrix order $\succeq$ is total for each of the pairs of matrices
$(\mb{A}', \mb{A}'')$ and $(\mb{B}', \mb{B}'')$ i.e., either
$\mb{A}' \succeq \mb{A}''$ (resp. $\mb{B}' \succeq \mb{B}''$) or
$\mb{A}'' \succ \mb{A}'$ (resp. $\mb{B}'' \succ \mb{B}'$). Then we
have $\mathrm{Tr}( \mb{M} + \mb{N}) \geq 0$ with $\mb{M} =
(\mb{A}''-\mb{A}') \left[ (\mb{I}+ \mb{A}')^{-1} -
(\mb{I}+\mb{A}'')^{-1}  \right]$, $\mb{N} = (\mb{B}''-\mb{B}')
\left[(\mb{I}+\mb{B}'+\mb{A}')^{-1} -
(\mb{I}+\mb{B}''+\mb{A}'')^{-1}\right]$.}
\end{lemma}
It can be shown (see Appendix \ref{appendix_1} for more details)
that the diagonally strict concavity condition writes in our setup
as $ p \mc{T}^{(1)} + \ol{p} \mc{T}^{(2)} > 0$ where $\forall s \in
\{1,2\}$, $\mc{T}^{(s)}$ is defined by $\mc{T}^{(s)} = \mathrm{Tr}(
\mb{M}^{(s)} + \mb{N}^{(s)}) $ where the matrices $\mb{M}^{(s)}$,
$\mb{N}^{(s)}$ have exactly the same structure as $\mb{M}$, $\mb{N}$
in the above Lemma. For example, if we consider two pairs of parameters
$(\alpha_1'$,$ \alpha_1'') \in \left(\mc{A}_1^{\mathrm{TPA}}
\right)^2 $ and $(\alpha_2'$,$ \alpha_2'') \in
\left(\mc{A}_2^{\mathrm{TPA}} \right)^2 $ such that either
$\alpha_1' \neq \alpha_1''$ or $\alpha_2' \neq \alpha_2''$
 as in
Theorem \ref{theo-2-rosen}, $\mc{T}^{(1)} $ can be obtained by using
the following matrices $\mb{A}' = \rho_1 \alpha_1' \mb{H}_1
\mb{H}_1^H $, $\mb{A}'' =
 \rho_1 \alpha_1'' \mb{H}_1 \mb{H}_1^H$, $\mb{B}' = \rho_2 \frac{1-\ol{p}\alpha_2'}{p} \mb{H}_2
 \mb{H}_2^H$, $\mb{B}''  =  \rho_2 \frac{1-\ol{p}\alpha_2''}{p}
\mb{H}_2\mb{H}_2^H$. The term $\mc{T}^{(2)}$ has a similar
form as $\mc{T}^{(1)}$ thus, applying Lemma \ref{lemma:diago} twice
and considering the special structure of the four matrices
($\mb{A}'$,
 $\mb{A}''$, $\mb{B}'$, $\mb{B}''$), one can prove that the term
$p \mc{T}^{(1)} + \ol{p} \mc{T}^{(2)}$ is strictly positive.
Therefore the unconditional uniqueness of the NE is guaranteed.
\end{proof}

\emph{Determination of the NE.} In order to determine the selfish PA
of the users at the NE, we now exploit the large system approach
derived in \cite{tulino-book-04} for single-user fading MIMO
channels. This will lead us to simple approximations of the utility
functions which are much easier to optimize. From now on, we assume
the asymptotic regime in terms of the number of antennas: $n_t
\longrightarrow \infty$, $n_r \longrightarrow \infty$,  and
$\ds{\lim_{n_t\rightarrow \infty,n_r \rightarrow \infty }
\frac{n_t}{n_r} = c < \infty}$.
In this asymptotic regime, references \cite{tulino-book-04,
sylverstein-jma-1995, tulino-it-2005} provide an equivalent of the
ergodic capacity of single-user MIMO channels, which corresponds
exactly to the situation seen by user $1$ (resp. $2$) when $S=1$
(resp. $S=2$); this gives directly the approximation of the rates
$R_1^{(1)}$ and $R_2^{(2)}$; see Eq. (\ref{eq:mimo-st-rates}). From
Eq. (\ref{eq:mimo-st-rates}) we also see that the rates $R_1^{(2)}$
and $R_2^{(1)}$ correspond to the difference between the sum-rate of
the equivalent $K n_t \times n_r$ virtual MIMO system and an $n_t
\times n_r$ single-user MIMO system, therefore the results of
\cite{tulino-book-04, sylverstein-jma-1995, tulino-it-2005} can also
be applied directly. The corresponding approximates can then be
easily checked to be:
 \begin{equation}
\begin{array}{lcl}
\tilde{R}_1^{(1)}(\alpha_1,\alpha_2) &=& \ds{\sum_{i=1}^{n_t}\log_2
\left[1+\eta \alpha_1P_1 d^{(\mathrm{T})}_1(i) \gamma_1\right]
}\ds{+ \sum_{j=1}^{n_r} \log_2 \left[1+ \eta d^{(\mathrm{R})}(j)
\delta_1 \right]
- n_t \eta \gamma_1 \delta_1  \log_2 e} \\
\tilde{R}_2^{(1)}(\alpha_1,\alpha_2) &=& \ds{ \sum_{i=1}^{n_t}\log_2
\left[1+2 \eta \alpha_1 P_1 d^{(\mathrm{T})}_2(i) \gamma_2\right]}
\ds{+ \sum_{i=1}^{n_t}\log_2 \left[1+2 \eta
\frac{1-\ol{p}\alpha_2}{p} P_2
d^{(\mathrm{T})}_2(i) \gamma_2\right]}\\
&& \ds{+ \sum_{j=1}^{n_r} \log_2 \left[1+ 2\eta d^{(\mathrm{R})}(j)
\delta_2\right] -}  4n_t \eta \gamma_2 \delta_2  \log_2 e -
\tilde{R}_1^{(1)}(\alpha_1,\alpha_2).
\end{array}
\end{equation}

where $\forall k \in \{1,2\}, d_k^{(\mathrm{T})}(i)$, $i \in
\{1,...,n_t\}$ are the eigenvalues of the transmit correlation
matrices $\mb{T}_k$ (see Eq. (\ref{eq:kronecker-model})),
$d^{(\mathrm{R})}(j)$, $j \in \{1,...,n_r\}$ , are the eigenvalues
of the receive correlation matrix $\mb{R}$ and the parameters
$\gamma_i$, $\delta_j$ are the unique solutions of the following
systems of $2-$degree equations:
 \begin{equation}
\label{alpha_sys1}
 \left\{
\begin{array}{ccl}
\gamma_1 & = & \ds{\frac{1}{n_t} \sum_{j=1}^{n_r} \frac{d^{(\mathrm{R})}(j)}{1 + \eta d^{(\mathrm{R})}(j) \delta_1}}\\
\delta_1 & =  & \ds{\frac{1}{n_t} \sum_{i=1}^{n_t} \frac{\alpha_1P_1 d^{(\mathrm{T})}_1(i)}{1 + \eta \alpha_1P_1 d^{(\mathrm{T})}_1(i)\gamma_1}}\\
\end{array}
\right.
\end{equation}
\begin{equation}
\label{alpha_sys2}
 \left\{
\begin{array}{ccl}
\gamma_2 & = & \ds{\frac{1}{2n_t} \sum_{j=1}^{n_r} \frac{d^{(\mathrm{R})}(j)}{1 +2 \eta d^{(\mathrm{R})}(j) \delta_2}}\\
\delta_2 & =  & \ds{\frac{1}{2n_t} \left[\sum_{i=1}^{n_t}
\frac{\alpha_1P_1 d^{(\mathrm{T})}_1(i)}{1 + 2\eta\alpha_1P_1
d^{(\mathrm{T})}_1(i) \gamma_2 }+ \sum_{i=1}^{n_t}
\frac{\frac{1-\ol{p}\alpha_2}{p}P_2 d^{(\mathrm{T})}_2(i)}{1
+ 2\eta\frac{1-\ol{p}\alpha_2}{p}P_2 d^{(\mathrm{T})}_2(i) \gamma_2 }\right]}.\\
\end{array}
\right.
\end{equation}

The approximate functions $\tilde{R}_1^{(2)}(\cdot,\cdot)$ and
$\tilde{R}_2^{(2)}(\cdot)$ can be obtained in a similar way and the
approximated utility of user $k\in\{1,2\}$ follows: $
\tilde{u}_k(\alpha_1,\alpha_2) = p
\tilde{R}_k^{(1)}(\alpha_1,\alpha_2) + \ol{p}
\tilde{R}_k^{(2)}(\alpha_1,\alpha_2)$ .
Now, in order to solve the constrained optimization problem, we
introduce the Lagrange multipliers
$(\lambda_{11},\lambda_{12},\lambda_{21},\lambda_{22}) \in [0, +
\infty )^4$ and define for $k\in \{1,2\}$ the function $
\mc{L}_k(\alpha_1,\alpha_2,\lambda_{k1},\lambda_{k2})  =
-\tilde{u}_k(\alpha_1,\alpha_2) +
\lambda_{k1}\left(\alpha_k-\frac{1}{p_k}\right)-\lambda_{k2}
\alpha_k$. The Kuhn-Tucker optimality conditions follow. Therefore,
the optimum selfish PAs,
$(\alpha_1^{\mathrm{NE}},\alpha_2^{\mathrm{NE}})$, can be obtained
by using a fixed-point method and an iterative algorithm, following
the same idea as in \cite{lasaulce-gamecomm-2007} for
non-coordinated MIMO MACs with single-user decoding. At this point
we have to make an important technical comment. Our proof for the
existence and uniqueness of the NE holds for the exact game. For the
approximated game, we need the approximated utilities to have the
same properties as their exact counterparts. It turns out that the
large system approximation of the ergodic mutual information can be
shown to have the desired properties \cite{dumont-arxiv-2007}. In
particular, the results of \cite{dumont-arxiv-2007} show that
 the approximated utilities are strictly concave and that if the iterative
 PA algorithm converges, it converges towards the global maximum.

\emph{Sum-rate efficiency of the NE.} Now, let us focus on the
sum-rate of the decentralized network and compare it with the
optimal sum-rate of its centralized counterpart. The centralized
network sum-rate, denoted by $R_{\mathrm{sum}}^{(\mathrm{C})}$, is
by definition obtained by jointly maximizing the sum-rate over all
the pairs of power fractions $(\alpha_1, \alpha_2) \in [0,1]^2$:
$\ds{R_{\mathrm{sum}}^{(\mathrm{C})} \triangleq \max_{(\alpha_1,
\alpha_2)} u_1(\alpha_1, \alpha_2) + u_2(\alpha_1, \alpha_2)}$.
Knowing that $\log|\cdot|$ is a concave function, one can easily
verify that the maximum is obtained for
$(\alpha_1^*,\alpha_2^*)=(1,1)$ and that $R_{\mathrm{sum}}^{(C)}=
\mathbb{E}\log
\left|\mb{I}+\rho_1\mb{H}_1\mb{H}_1^H+\rho_2\mb{H}_2\mb{H}_2^H\right|$.
 As the optimum precoding matrices are proportional to the identity
matrix, it can be checked that the network sum-rate at the NE
(denoted by $R_{\mathrm{sum}}^{\mathrm{NE}}$) is equal to the
centralized network sum-rate for $p=0$ and $p=1$:
$R_{\mathrm{sum}}^{\mathrm{NE}}(0) =
R_{\mathrm{sum}}^{\mathrm{NE}}(1)= R_{\mathrm{sum}}^{(\mathrm{C})}$.
Indeed, let us consider that $p=1$. In this case, user 1 is always
decoded in the second place ($\mathrm{Pr}[S=1]=1$). This means that
there is no temporal power allocation game here and each user
always allocates all of his available power for the case where
$S=1$: $(\alpha_1^{\mathrm{NE}}, \alpha_2^{\mathrm{NE}})=(1,0)$.
Replacing in Eq. (4) the corresponding correlation matrices:
$\mb{Q}_1^{(1)}=\mb{I}_{n_t}$, $\mb{Q}_2^{(1)}=\mb{I}_{n_t}$ and
$\mb{Q}_1^{(2)}=\mb{O}_{n_t}$ (the square zero matrix),
$\mb{Q}_2^{(2)}=\mb{O}_{n_t}$ we obtain that
$R_{\mathrm{sum}}^{\mathrm{NE}}(1)=R_{\mathrm{sum}}^{(\mathrm{C})}$.

In the \emph{high SNR regime}, where $\eta \rightarrow \infty$, we
obtain from (\ref{alpha_sys1}),(\ref{alpha_sys2}) that $\eta
\delta_1 \rightarrow \frac{1}{\gamma_1}$, $\eta \delta_2 \rightarrow
\frac{1}{2\gamma_2}$ and thus $\gamma_1$ and $\gamma_2$ are the
unique solutions of the following equations:
$\frac{1}{n_t}\sum_{j=1}^{n_r}\frac{d^{(\mathrm{R})}(j)}{\gamma_1+d^{(\mathrm{R})}(j)}=1$,
 $\frac{1}{2n_t}\sum_{j=1}^{n_r}\frac{d^{(\mathrm{R})}(j)}{\gamma_2+d^{(\mathrm{R})}(j)}=1$.
The approximated utilities become:
 \begin{equation}
\begin{array}{lcl}
\tilde{R}_1^{(1)}(\alpha_1,\alpha_2) &=& \ds{\sum_{i=1}^{n_t}\log_2
\left[1+\eta \alpha_1P_1 d^{(\mathrm{T})}_1(i) \gamma_1\right] }
\ds{+ \sum_{j=1}^{n_r} \log_2 \left[1+ \frac{d^{(\mathrm{R})}(j)}{\gamma_1} \right] - n_t  \log_2 e} \\
\tilde{R}_2^{(1)}(\alpha_1,\alpha_2) &=& \ds{ \sum_{i=1}^{n_t}\log_2
\left[1+2 \eta \alpha_1 P_1 d^{(\mathrm{T})}_1(i) \gamma_2\right]}
\ds{+ \sum_{i=1}^{n_t}\log_2 \left[1+2 \eta
\frac{1-\ol{p}\alpha_2}{p} P_2 d^{(\mathrm{T})}_2(i)
\gamma_2\right]}
\ds{+ \sum_{j=1}^{n_r} \log_2 \left[1+ \frac{d^{(\mathrm{R})}(j)}{\gamma_2} \right] } \\
& &- 2n_t  \log_2 e- \tilde{R}_1^{(1)}(\alpha_1,\alpha_2).
\end{array}
\end{equation}
By setting the derivatives of $\tilde{u}_1(\cdot,\cdot)$ w.r.t.
$\alpha_1$ and $\tilde{u}_2(\cdot,\cdot)$ w.r.t. $\alpha_2$ to zero,
we obtain that, for each user, the PA at the NE is the uniform PA
$(\alpha_1^{\mathrm{NE}},\alpha_2^{\mathrm{NE}})=(1,1)$, regardless
of the distribution of the coordination signal $p \in [0,1]$.
Therefore, at the equilibrium, we have that
\begin{eqnarray}
R_{\mathrm{sum}}^{\mathrm{NE}}(p) & = &
pR_1^{(1)}(\alpha_1^{\mathrm{NE}},\alpha_2^{\mathrm{NE}}) +
\ol{p}R_1^{(2)}(\alpha_1^{\mathrm{NE}},\alpha_2^{\mathrm{NE}}) +p
R_2^{(1)}(\alpha_1^{\mathrm{NE}},\alpha_2^{\mathrm{NE}}) +\ol{p}R_2^{(2)}(\alpha_1^{\mathrm{NE}},\alpha_2^{\mathrm{NE}}) \nonumber\\
& = & p \mathbb{E} \log | \mb{I} + \rho_1 \mb{H}_1 \mb{H}_1^H | +
\ol{p}\mathbb{E} \log | \mb{I} + \rho_1 \mb{H}_1 \mb{H}_1^H + \rho_2
\mb{H}_2 \mb{H}_2^H|-\ol{p}\mathbb{E} \log | \mb{I} +  \rho_2
\mb{H}_2 \mb{H}_2^H|  \\
& & + p \mathbb{E} \log | \mb{I} + \rho_1 \mb{H}_1 \mb{H}_1^H +
\rho_2 \mb{H}_2 \mb{H}_2^H|-p\mathbb{E} \log | \mb{I} + \rho_1
\mb{H}_1 \mb{H}_1^H | +\ol{p}\mathbb{E} \log | \mb{I} + \rho_2 \mb{H}_2 \mb{H}_2^H|  \nonumber \\
&=& R_{\mathrm{sum}}^{(\mathrm{C})} \nonumber.
\end{eqnarray}
Knowing that the uniform spatial PA is optimal in the high SNR
regime \cite{telatar-ett-1999, lasaulce-gamecomm-2007}, the
centralized network sum-rate coincides with the sum-capacity of the
centralized MAC channel,
$R_{\mathrm{sum}}^{(\mathrm{C})}=C_{\mathrm{sum}}$.

In the \emph{low SNR regime}, where $\eta \rightarrow 0$, we obtain
from (\ref{alpha_sys1}), (\ref{alpha_sys2}) that $\eta \delta_1
\rightarrow 0$, $\eta \delta_2 \rightarrow 0$ and thus $\gamma_1=
\frac{1}{n_t}\sum_{j=1}^{n_r}d^{(\mathrm{R})}(j)$, $\gamma_2=
\frac{1}{2n_t}\sum_{j=1}^{n_r}d^{(\mathrm{R})}(j)$ . Approximating
$\ln(1+x)\approx x$ for $x<<1$, the achievable rates become:
\begin{equation}
\begin{array}{lcl}
\tilde{R}_1^{(1)}(\alpha_1) &=& \ds{\frac{1}{n_t} \eta P_1 \alpha_1 \sum_{j=1}^{n_r}d^{(\mathrm{R})}(j)\sum_{i=1}^{n_t}d^{(\mathrm{T})}_1(i) \log_2 e} \\
\tilde{R}_2^{(1)}(\alpha_1,\alpha_2) &=& \ds{\frac{1}{n_t}
\eta\frac{1-\ol{p}\alpha_2}{p}P_2\sum_{j=1}^{n_r}d^{(\mathrm{R})}(j)\sum_{i=1}^{n_t}d^{(\mathrm{T})}_1(i)
\log_2 e}
\end{array}.
\end{equation}
We see that the utilities $\tilde{u}_k(\alpha_1,\alpha_2)
=\frac{1}{n_t} \eta P_k
\sum_{j=1}^{n_r}d^{(\mathrm{R})}(j)\sum_{i=1}^{n_t}d^{(\mathrm{T})}_k(i)
\log_2 e$ converge and the network sum-rate at the NE coincides here
again with the centralized network sum-rate:

$R_{\mathrm{sum}}^{(\mathrm{C})}= \frac{1}{n_t}
\sum_{j=1}^{n_r}d^{(\mathrm{R})}(j)\left(\eta
P_1\sum_{i=1}^{n_t}d^{(\mathrm{T})}_1(i) + \eta
P_2\sum_{i=1}^{n_t}d^{(\mathrm{T})}_2(i) \right)\log_2 e$. In this
case also, the price of anarchy \cite{papadimitriou-stoc-2001} is
minimal for any distribution of the coordination signal.

To sum up we have seen that there is no loss of optimality in terms
of sum-rate by decentralizing the PA procedure in at least four
special cases: 1) $p=0$; 2) $p=1$; 3) when $\eta \rightarrow \infty$
for any $p \in [0,1]$; 4) when $\eta \rightarrow 0$ for any $p \in
[0,1]$. Additionally, in case 3), since there is no loss by imposing
the spatially uniform PA \cite{telatar-ett-1999,
lasaulce-gamecomm-2007}, the centralized (and cooperative) MAC
sum-capacity is achieved. If we further assume that there is no
correlation among the transmit antennas, $\mb{T}_k=\mb{I}$, the
uniform spatial PA is optimal \cite{telatar-ett-1999} for any
$\eta$. Thus, the centralized sum-rate is always identical to the
sum-capacity of the centralized MAC channel,
$R_{\mathrm{sum}}^{(\mathrm{C})} = C_{\mathrm{sum}}$. This means
that if the BS chooses to use a completely unfair SIC-based decoding
scheme, the selfish behavior of the users will always lead to
achieving the centralized sum-capacity. This result is in agreement
with \cite{lai-it-2008}, where the authors have proposed a
water-filling game for the fast fading SISO MAC (assuming perfect
CSIT and CSIR) and shown that the equilibrium sum-rate is equal to
the maximum sum-rate point of the capacity region. However, as opposed to
 the SISO MAC with the proposed coordination mechanism
\cite{belmega-wnc3-2008}, the decentralized MIMO MAC with
coordination does not achieve the sum-rate of the equivalent virtual
MIMO network for any value of $p$ and for an arbitrary noise level
at the BS. In particular, the fair choice $p = \frac{1}{2}$ is not
optimal. We will quantify the corresponding performance gap through
simulation results. Furthermore, in the low and high SNR regimes,
the centralized sum-capacity is also achieved for any value of $p$.
The consequence of these results is that any binary coordination
signal can be used without loss of global optimality.

\section{Spatial power allocation game}
\label{sec:spatial-PA-game} In this section, we assume that the users are free to share their
transmit power between their antennas but for each realization of
the coordination signal the transmit power is constrained by Eq.
(6). In other words
 we assume that the users cannot distribute their power
over time: they cannot decide the amount of power they dedicate to a
given realization of the coordination signal. As a consequence
of this power constraint (Eq. (6)), the two precoding matrices that each
user needs to choose can be optimized independently and each of them
does not depend on $p$. The strategy
set of user $k$ is in the spatial PA (SPA) game is:
\begin{equation}
\begin{array}{lcl}  \mc{A}_k^{\mathrm{SPA}} = \left\{
\mb{Q}_k=(\mb{Q}_k^{(1)},\mb{Q}_k^{(2)})\right. & | & \mb{Q}_k^{(1)}
\succeq
0, \mb{Q}_k^{(1)}=\mb{Q}_k^{(1)H}, \mathrm{Tr}(\mb{Q}_k^{(1)}) \leq n_t P_k \\
& & \left. \mb{Q}_k^{(2)} \succeq 0, \mb{Q}_k^{(2)}=\mb{Q}_k^{(2)H},
\mathrm{Tr}(\mb{Q}_k^{(2)}) \leq n_t P_k   \right\}.
\end{array}
\end{equation}

\begin{theorem}[Existence and uniqueness of an NE in Game 2]
\emph{The SPA game defined by the set of players $\mc{K}= \{1,2\}$,
the strategy sets $\mc{A}_k^{(\mathrm{SPA})}$ and utilities
$u_k(\alpha_k,\alpha_{-k})$ given by Eq. (\ref{eq:utility-mimo}),
has a unique NE.}
\end{theorem}

\begin{proof}
The main feature of the game under the aforementioned power
constraint is that there exists a unique NE in each sub-game defined
by the realization of the coordination signal. The proof is much
simpler than that of the time PA problem since the use of Rosen's
Theorem \cite{rosen-eco-1965} is not required. Without loss of
generality assume that $S=1$. Whatever the strategy of user 2, user
1 sees no interference. Therefore he can choose $\mb{Q}_1^{(1)}$
independently of user 2. Because $R_1^{(1)}(\mb{Q}_1^{(1)},
\mb{Q}_2^{(1)})$ is a strictly concave function to be maximized over
a convex set, there is a unique optimum strategy for user 1. As we
assume a game with complete information and rational users, user 2
knows the utility of user 1 and thus the precoding matrix he will choose. The same concavity argument can be used
for $R_2^{(1)}(\mb{Q}_1^{(1)}, \mb{Q}_2^{(1)})$ and therefore
guarantees that user 2 employs a unique precoding matrix.
\end{proof}

\emph{Determination of the NE.} In order to find the optimum
covariance matrices, we proceed in the same way as described in
\cite{lasaulce-gamecomm-2007}. First we focus on the optimum
eigenvectors and then we determine the optimum eigenvalues by
approximating the utility functions under the large system
assumption. In order to determine the optimum eigenvectors, the
proof in \cite{soysal-tcsubmitted-2007} can be applied in our
context to assert that there is no loss of optimality by restricting
the search for the optimum covariance matrix when imposing the
structure $\mb{Q}_k^{(s)}=\mb{U}_k\mb{P}_k^{(s)}\mb{U}_k^H$, where
$\mb{U}_k$ is a unitary matrix coming from the spectral
decomposition of transmit correlation matrix
$\mb{T}_k=\mb{U}_k\mb{D}_k\mb{U}_k^H$ defined in Eq.
(\ref{eq:kronecker-model}) and the diagonal matrix $\mb{P}_k^{(s)}
=\text{ Diag}(P_k^{(s)}(1), ..., P_k^{(s)}(n_t))$ represents the
powers user $k$ allocates to the different eigenvectors. As a
consequence, we can exploit once again the results of
\cite{tulino-book-04,sylverstein-jma-1995,tulino-it-2005} assuming
the asymptotic
 regime in terms of the number of antennas.
The new approximated rates are:
 \begin{equation}
\label{eq:mimo-s-approx}
\begin{array}{lcl}
\tilde{R}_1^{(1)}(\mb{P}_1^{(1)}) &=& \ds{\sum_{i=1}^{n_t}\log_2 \left[1+\eta P_1^{(1)}(i)d^{(\mathrm{T})}_1(i) \gamma_1\right] }\ds{+ \sum_{j=1}^{n_r} \log_2 \left[1+ \eta d^{(\mathrm{R})}(j) \delta_1 \right] - n_t \eta \gamma_1 \delta_1  \log_2 e} \\
\tilde{R}_2^{(1)}(\mb{P}_1^{(1)},\mb{P}_2^{(1)}) &=&
\ds{\sum_{\ell=1}^2 \sum_{i=1}^{n_t}\log_2 \left[1+2 \eta P_{\ell}^{(1)}(i)d^{(\mathrm{T})}_{\ell}(i) \gamma_2\right]}  \ds{+ \sum_{j=1}^{n_r} \log_2 \left[1+ 2\eta d^{(\mathrm{R})}(j) \delta_2\right] - 4n_t \eta \gamma_2 \delta_2  \log_2 e}- \\
& & - \tilde{R}_1^{(1)}(\mb{P}_1^{(1)})
\end{array}
\end{equation}
where $\forall k \in \{1,2\}, d_k^{(\mathrm{T})}(i)$, $i \in
\{1,...,n_t\}$ are always the eigenvalues of the transmit
correlation matrices $\mb{T}_k$, $d^{(\mathrm{R})}(j)$, $j \in
\{1,...,n_r\}$ , are the eigenvalues of the receive correlation
matrix $\mb{R}$ and the parameters $\gamma_i$, $\delta_j$ are the
unique solutions of the following systems of equations:
 \begin{equation}
\label{eq:mimo-s-approx_gamma1}
 \left\{
\begin{array}{ccl}
\gamma_1 & = & \ds{\frac{1}{n_t} \sum_{j=1}^{n_r} \frac{d^{(\mathrm{R})}(j)}{1 + \eta d^{(\mathrm{R})}(j) \delta_1}}\\
\delta_1 & =  & \ds{\frac{1}{n_t} \sum_{i=1}^{n_t} \frac{P_1^{(1)}(i) d^{(\mathrm{T})}_1(i)}{1 + \eta P_1^{(1)}(i) d^{(\mathrm{T})}_1(i)\gamma_1}}\\
\end{array}
\right.
\end{equation}
\begin{equation}
\label{eq:mimo-s-approx_gamma2}
 \left\{
\begin{array}{ccl}
\gamma_2 & = & \ds{\frac{1}{2n_t} \sum_{j=1}^{n_r} \frac{d^{(\mathrm{R})}(j)}{1 +2 \eta d^{(\mathrm{R})}(j) \delta_2}}\\
\delta_2 & =  & \ds{\frac{1}{2n_t}  \sum_{l=1}^2 \sum_{i=1}^{n_t} \frac{P_l^{(1)}(i)
d^{(\mathrm{T})}_l(i)}{1 + 2\eta P_l^{(1)}(i) d^{(\mathrm{T})}_l(i)1 \gamma_2 }}.\\
\end{array}
\right.
\end{equation}
Then, optimizing the approximated rates $\tilde{R}_k^{(1)} (\cdot)$
w.r.t. $P_k^{(1)}(i)$ leads to the following water-filling
equations:
\begin{equation}
\label{waterfill1}
\forall k \in \{1,2\}, \ P_k^{(1),\mathrm{NE}}(i) = \left[ \frac{1}{
\ln 2 \lambda_k^{(1)}} - \frac{1}{\eta d^{(\mathrm{T})}_k(i)
\gamma_k}\right]^+
\end{equation}
where $\lambda_k^{(1)} \geq 0$, $k \in \{1,2\}$, are the Lagrangian
multipliers tuned in order to meet the power constraints given in
(\ref{eq:constraint-spatial-pa}): $\sum_{i=1}^{n_t}
P_k^{(1),\mathrm{NE}}(i) = n_t P_k$. We use the same iterative PA
algorithm as the one described in \cite{lasaulce-gamecomm-2007}.
Under the large systems assumption, in this game also, the
approximated utilities have the same properties as the exact
utilities.

\emph{Sum-rate efficiency of the NE.} Unlike the temporal PA game,
we have not assumed a particular structure for the precoding
matrices and thus the centralized solution coincides with the
sum-capacity of the virtual MIMO network,
$R_{\mathrm{sum}}^{(\mathrm{C})}=C_{\mathrm{sum}}$. Another
important point to notice here is that the equilibrium precoding
matrices do not depend on $p$. This considerably simplifies the BS's
choice for the sum-rate optimal value for $p$. Indeed, as we have already
mentioned, the precoding matrices do no depend on $p$ and therefore
the sum-rate $R_{\mathrm{sum}}(p)$ is merely a linear function of
$p$: $R_{\mathrm{sum}}^{\mathrm{NE}}(p) = a p + b$ where
\begin{equation}
\label{ab}
\begin{array}{ccl}
a &=& \mathbb{E} \log |\mb{I} + \eta
\mb{H}_1\mb{Q}_1^{(1),\mathrm{NE}}
 \mb{H}_1^H + \eta \mb{H}_2\mb{Q}_2^{(1),\mathrm{NE}} \mb{H}_2^H  | -  \mathbb{E} \log |\mb{I} + \eta \mb{H}_1\mb{Q}_1^{(2),\mathrm{NE}}
  \mb{H}_1^H + \eta \mb{H}_2\mb{Q}_2^{(2),\mathrm{NE}} \mb{H}_2^H  | \\
b &=& \mathbb{E} \log |\mb{I} + \eta
\mb{H}_1\mb{Q}_1^{(2),\mathrm{NE}}
  \mb{H}_1^H + \eta \mb{H}_2\mb{Q}_2^{(2),\mathrm{NE}} \mb{H}_2^H  |.
\end{array}
\end{equation}
Depending on the sign of $a$, if the BS wants to maximize the
sum-rate, it will choose either $p=0$ or $p=1$. If it wants a fair
game it will choose $p = \frac{1}{2}$ and accept a certain loss of
global optimality. Note that even for $p \in \{0,1\}$ the
sum-capacity is not reached in general: this is because the matrix
$\mb{Q}_1^{(1),\mathrm{NE}}$ (resp. $\mb{Q}_2^{(2),\mathrm{NE}}$)
does not coincide with the first (resp. second) component of the
pair of precoding matrices that maximizes the (strictly concave)
network sum-rate. However, as we did for the temporal PA game, in
the low and high SNR regimes one can show that the decentralized
MIMO MAC has the same performance (w.r.t. the sum-rate) as its
equivalent $Kn_t \times n_r$ virtual MIMO network.

\section{SIMULATION EXAMPLES}
\label{sec:simulation-results} 

All the results will be provided by assuming the asymptotic regime
in the numbers of antennas. We know, from many contributions (see
e.g.,
\cite{lasaulce-gamecomm-2007,dumont-arxiv-2007,biglieri-issta-02,dumont-globecom-2006})
that large-system approximates of ergodic rates are accurate even
for relatively small systems. We also assume that $\mb{R}=\mb{I}$.

\begin{figure}[!b]
  \begin{center}
    \includegraphics[scale=0.60]{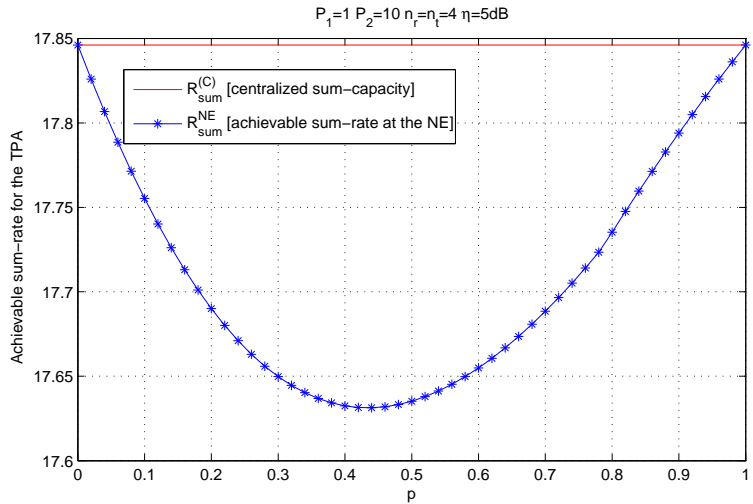}
  \end{center}
  \caption{ \footnotesize Temporal PA game. Achievable network sum-rate versus $p$
  for $P_1=1$, $P_2=10$, $n_r=n_t=4$, $\eta=5$ dB. The sum-capacity of fading MIMO MACs is reached for both unfair SIC decoding schemes ($p^{*}_1=0$ and $p^{*}_2=1$) and is very close to
  this upper bound for any
  distribution of the coordination signal, $\forall p \in (0,1)$.}
  \label{fig1}
\end{figure}

For the TPA problem, we look at the case where there is no transmit
correlation, $\mb{T}_k=\mb{I}$. We have seen that the performance of
decentralized MAC depends on the rule of the game i.e., the value of
$p$. This is exactly what Fig. \ref{fig1} depicts for the following
scenario: $P_1=1$, $P_2 = 10$, $\eta = 5$ dB, $n_t = n_r =4$. First,
we see that the MAC sum-rate is a convex function of $p$ and the
maximum of $R_{\mathrm{sum}}^{\mathrm{NE}}(p)$ is reached for
$p\in\{0,1\}$. In these points, which correspond to the most unfair
decoding schemes (either user 1 or 2 is always decoded first) the
centralized sum-capacity of the MAC is achieved. One important
observation to be made is that the minimum and maximum only differ
by about $1 \%$. Many other simulations have confirmed this
observation. This shows that whatever the value of $p$, the gap
between the sum-rate of a decentralized MIMO MAC with selfish users
and the sum-capacity of the equivalent cooperative MAC (virtual MIMO
network) is in fact very small. Now, we want to evaluate the
benefits brought by using a SIC instead of single-user decoding
\cite{lasaulce-gamecomm-2007}. For the scenario where $P_1=P$, $P_2=10P$ with $P \in [0,20]$, $n_r=n_t=4$ and $\eta=5$ dB, Fig. 2 shows the
     achievable network sum-rate at the NE versus
    the available power at the first transmitter $P$. For the SUD scheme, the
users are decoded simultaneously at the receiver. In this case both
users see all the interference coming from the others. We see that
the SIC scheme performs much better than the proposed SUD scheme,
regardless of the distribution of the coordination signal: this
comparison makes sense especially for the point $p= \frac{1}{2}$
since both decoding schemes are fair.

From now on, we consider the
SPA problem. In this case we assume an exponential correlation
profile for $\mb{T}_k$ such that $\mb{T}_k(i,j)= t_k^{|i-j|}$ (note that
$\mathrm{Tr}(\mb{T}_k)=n_t$), where $0 \leq t_k \leq 1$ is the
corresponding correlation coefficient \cite{chiani-it-2003,
skupch-pwc-2005}. We already know that the sum-rate is a linear
function of $p$ and therefore is maximized when either $p=0$ or $p=1$.
It turns out that this slope has a small value. Furthermore, it has been observed to be even
 0 for a symmetric MAC, i.e., $P_1=P_2$ and $t_1=t_2$. These observations have been confirmed by
  many simulations. In Fig. 3 we have plotted the sum-rate achieved by varying $p$ for the
   scenario: $P_1=5$, $P_2=50$, $\eta=3$ dB, $n_t=n_r=4$, $t_1=0.4$, $t_2=0.3$. Even in this scenario, which was thought to be
a bad case in terms of sub-optimality, the sum-rate is not far from
the sum-capacity of the centralized MAC. For the same scenario, we have plotted in Fig. 4 the achievable
rate region and compared it to that obtained with SUD. We observe
that in large MIMO
MAC channels, the capacity region comprises a full cooperation
segment (approximately) just like SISO MAC channels. The
coordination signal allows one to move along an almost straight
line, corresponding to a relatively large range of rates.

\section{CONCLUSION}
\label{sec:conclusions}

We have provided complete proofs for the existence and uniqueness
of an NE in fast fading MIMO MACs with CSIR and CDIT where the
transmission rate is chosen as user utility. By exploiting random
matrix theory, we have also provided the corresponding optimum
selfish PA policies. We have seen that the BS can, through a single
parameter (i.e., $p \in [0,1]$, which represents the distribution of
the coordination signal), force the system to operate at many
different points that correspond to a relatively large range of
achievable transmission rate pairs. We know, from
\cite{belmega-wnc3-2008, lai-it-2008} that for Gaussian MACs with
single antenna terminals, this set of rate pairs corresponds to the
full cooperation segment of the centralized MAC. Said otherwise a
\emph{decentralized} Gaussian SISO MAC with coordination achieves
the same rate pairs as a MAC with full cooperation or \emph{virtual
MIMO system}. The goal here was to know to what extent this key
result is valid for fading MAC with multi-antenna terminals. It
turns out this is almost true in the MIMO setting. In the cases of
interest, where the power is optimally allocated either over space or
time, the performance gap is relatively small even though the
proposed coordination mechanism was a priori sub-optimal since it
does take into account the channel realizations (known to the
receiver). Interestingly in large MIMO MACs, the capacity region
comprises a full cooperation segment just like SISO MACs. The
coordination signal precisely allows one to move along the
corresponding (almost) straight line. This shows the relevance of
large systems in decentralized networks since they allow to
determine the capacity region of certain systems whereas it is
unknown in the finite setting. Furthermore, they induce an averaging effect,
which makes the users' behavior predictable. Indeed, in large MIMO
MACs the knowledge of the CSIT does not improve the performance
w.r.t. the case with CDIT. To conclude we review some extensions of
this work which we have suggested throughout it. It would be
interesting to study the case of the decentralized space-time PA,
which, in particular, would require the generalization of Lemma
\ref{lemma:diago} to arbitrary positive matrices and exploitation of
some results in \cite{girko-book-2001}. A second useful extension
would be to evaluate the impact of a non-perfect SIC on the PA
problem. At last, we will mention that it would be useful to
evaluate analytically or bounding the price of anarchy of the NE,
which
 would require to find a bounding technique different from that used
 for non-atomic games \cite{roughgarden-jcss-2003, correa-ipco-2005,
maille-workingpaper-2007}.

\appendices
\appendices
\section{}
\label{appendix_1} We want to prove that the diagonally strict
concavity condition is met for the time PA problem i.e., for all
$(\alpha_1'$,$ \alpha_1'') \in \left(\mc{A}_1^{\mathrm{TPA}}
\right)^2 $ and $(\alpha_2'$,$ \alpha_2'') \in
\left(\mc{A}_2^{\mathrm{TPA}} \right)^2 $ such that either
$\alpha_1' \neq \alpha_1''$ or $\alpha_2' \neq \alpha_2''$ we want
to prove that:
\begin{equation}
\mc{C} = (\alpha_1''-\alpha_1') \left[  \frac{\partial R_1}{\partial
\alpha_1} (\alpha_1',\alpha_2') - \frac{\partial R_1}{\partial
\alpha_1} (\alpha_1'',\alpha_2'')\right] + (\alpha_2''-\alpha_2')
\left[  \frac{\partial R_2}{\partial \alpha_2} (\alpha_1',\alpha_2')
- \frac{\partial R_2}{\partial \alpha_2}
(\alpha_1'',\alpha_2'')\right] >0.
\end{equation}

We can write $\mc{C}= p\mc{T}^{(1)}+\ol{p}\mc{T}^{(2)}$ where for
all $s\in \{1,2\}$:
\begin{equation}
\mc{T}^{(s)}=(\alpha_1''-\alpha_1') \left[  \frac{\partial
R_1^{(s)}}{\partial \alpha_1} (\alpha_1',\alpha_2') - \frac{\partial
R_1^{(s)}}{\partial \alpha_1} (\alpha_1'',\alpha_2'')\right]
+(\alpha_2''-\alpha_2') \left[  \frac{\partial R_2^{(s)}}{\partial
\alpha_2} (\alpha_1',\alpha_2') - \frac{\partial R_2^{(s)}}{\partial
\alpha_2} (\alpha_1'',\alpha_2'')\right]
\end{equation}
By expanding $\mc{T}^{(1)}$ we have

\begin{equation}
\begin{array}{lcl}
\mc{T}^{(1)}&=&(\alpha_1'' -\alpha_1') \mathbb{E} \mathrm{Tr}
\left\{ [ (\mb{I}+\rho_1 \alpha_1' \mb{H}_1 \mb{H}_1^H)^{-1} -
(\mb{I}+\rho_1 \alpha_1' \mb{H}_1
\mb{H}_1^H)^{-1}]\rho_1 \mb{H}_1 \mb{H}_1^H \right\} \\
& & + (\alpha_2'' -\alpha_2') \mathbb{E} \mathrm{Tr} \left\{
(\mb{I}+\rho_1 \alpha_1' \mb{H}_1 \mb{H}_1^H  +\frac{1-\ol{p}
\alpha_2'}{p} \rho_2 \mb{H}_2 \mb{H}_2^H)^{-1}
\rho_2\frac{-\ol{p}}{p} \mb{H}_2 \mb{H}_2^H \right.
\\
& & -  \left. (\mb{I}+\rho_1 \alpha_1'' \mb{H}_1 \mb{H}_1^H
+\frac{1-\ol{p} \alpha_2''}{p} \rho_2 \mb{H}_2 \mb{H}_2^H)^{-1}
\rho_2 \frac{-\ol{p}}{p}\mb{H}_2
\mb{H}_2^H \right\}\\
&=&\mathbb{E} \mathrm{Tr} \left\{(\mb{A}''-\mb{A}') [ (\mb{I}+
\mb{A}')^{-1} - (\mb{I}+\mb{A}'')^{-1}  ]   +  (\mb{B}''- \mb{B}')
[(\mb{I}+\mb{B}'+\mb{A}')^{-1} -
(\mb{I}+\mb{B}''+\mb{A}'')^{-1}]\right\},
\end{array}
\end{equation}
 where $\mb{A}'  =  \rho_1 \alpha_1' \mb{H}_1 \mb{H}_1^H$, $\mb{A}'' =  \rho_1 \alpha_1'' \mb{H}_1
 \mb{H}_1^H$, $\mb{B}' =  \rho_2 \frac{1-\ol{p}\alpha_2'}{p} \mb{H}_2 \mb{H}_2^H
 $, $\mb{B}'' =  \rho_2 \frac{1-\ol{p}\alpha_2''}{p}
 \mb{H}_2\mb{H}_2^H$. We observe that the matrices $\mb{A}'$, $\mb{A}''$, $\mb{B}'$ and $\mb{B}''$ verify the
 assumptions of Lemma \ref{lemma:diago}. First, they are Hermitian and
 non-negative. Second, as they write as $\mb{A}' = a' \mb{H}_1 \mb{H}_1^H$, $\mb{A}'' = a'' \mb{H}_1 \mb{H}_1^H$,
 $\mb{B}' = b' \mb{H}_2 \mb{H}_2^H$ and $\mb{B}'' = b'' \mb{H}_2 \mb{H}_2^H$,
 we also see that the matrix order $\succeq$ is total for each of the pairs of
 matrices $(\mb{A}', \mb{A}'')$ and $(\mb{B}', \mb{B}'')$. This directly follows from the fact
 that the scalar order $\geq$
 is total, which implies that either $a''\geq a'$ or $a'' \leq a'$
 and either $b'' \geq b'$ or $b'' \leq b'$. By considering the particular structure of the four
 matrices and applying Lemma \ref{lemma:diago}, it is
 straightforward to see that the term $\mc{T}^{(1)}$
 is strictly positive, $\mc{T}^{(1)}> 0$. In a similar way we can prove that $\mc{T}^{(2)}
> 0$ and thus the diagonally strict concavity condition is met:
$\mc{C}>0$.

\section{}
\label{appendix_2}

Proving Lemma \ref{lemma:diago} amounts to showing that
\begin{equation}
 \mc{T}=\mathrm{Tr}
\left\{(\mb{A}-\mb{B})(\mb{B}^{-1}-\mb{A}^{-1}) +
(\mb{C}-\mb{D})[(\mb{B} + \mb{D})^{-1}-(\mb{A}+\mb{C})^{-1}]\right\}
> 0
\end{equation}
where the matrices $\mb{A} = \mb{I} + \mb{A}'' $, $\mb{B} = \mb{I} +
\mb{A}'$, $\mb{C} = \mb{B}''$ and $\mb{D} = \mb{B}'$ have been
introduced for more clarity. 
 Since the matrix
order $\succeq$ is total for $\mb{A}$ and $\mb{B}$, and $\mb{C}$ and
$\mb{D}$ it suffices to prove that $\mc{T} > 0$ for the four
following cases: (1) $\mb{A} \succeq \mb{B}$ and  $\mb{C} \succeq
\mb{D}$; (2) $\mb{A} \preceq \mb{B}$ and  $\mb{C} \preceq \mb{D}$;
(3) $\mb{A} \succeq \mb{B}$ and  $\mb{C} \preceq \mb{D}$; (4)
$\mb{A} \preceq \mb{B}$ and $\mb{C} \succeq \mb{D}$.

\textbf{Case (1): $\mb{A} \succeq \mb{B}$ and  $\mb{C} \succeq
\mb{D}$.} To prove the desired result in this case we use the
following lemma.
\begin{lemma}
\label{trace_prod} \emph{If $\mb{M}$ is a Hermitian and non-negative
($\mb{M}=\mb{M}^H \succeq \mb{0}$) and $\mb{N}$ is non-negative
($\mb{N} \succeq 0$) but not necessarily Hermitian, then
$\mathrm{Tr} (\mb{M}\mb{N}) \geq 0.$}
\end{lemma}
\begin{proof}
We write $\mathrm{Tr}(\mb{M}\mb{N})=
\mathrm{Tr}(\mb{M}^{1/2}\mb{N}\mb{M}^{1/2}) \geq 0$.
We have used the fact that $\mb{M}$ is a Hermitian non-negative
matrix to write $\mb{M}=\mb{M}^{1/2}\mb{M}^{1/2}$. Knowing that
$\mb{N}$ is a non-negative matrix one can easily check that
$\mb{M}^{1/2}\mb{N}\mb{M}^{1/2}$ is also a non-negative matrix and
thus the trace (sum of the non-negative eigenvalues) is
non-negative.
\end{proof}
The quantity $\mc{T}$ writes as $ \mc{T} =
\mathrm{Tr}(\mb{M}_1\mb{N}_1) + \mathrm{Tr}(\mb{M}_2\mb{N}_2)$ where
$\mb{M}_1= \mb{A}-\mb{B}$, $\mb{N}_1= \mb{B}^{-1}-\mb{A}^{-1}$,
$\mb{M}_2= \mb{C}-\mb{D}$ and $\mb{N}_2= (\mb{B} +
\mb{D})^{-1}-(\mb{A}+\mb{C})^{-1}$. Clearly these four matrices are
Hermitian. Since by assumption $\mb{M}_1 \succeq 0$ and $\mb{M}_2
\succeq 0$ we only need to verify that $\mb{N}_1 \succeq 0$ and
$\mb{N}_2 \succeq 0$ to be able to apply Lemma \ref{trace_prod} to
$\mc{T}$. The matrix $\mb{N}_1$ is non-negative because for any pair
of invertible matrices $(\mb{X}, \mb{Y})$: $\mb{X}\succeq \mb{Y}
\Leftrightarrow \mb{Y}^{-1}\succeq \mb{X}^{-1}$ (see e.g.,
\cite{horn-book-2007}). The same result applies to $\mb{N}_2$ since
by assumption $ \mb{A}+\mb{C} \succeq \mb{B}+\mb{D}$. Using lemma
\ref{trace_prod} concludes the proof.

\textbf{Case (3): $\mb{A} \succeq \mb{B}$ and  $\mb{C} \preceq
\mb{D}$.} To treat this case we first prove the following auxiliary
Lemma.
\begin{lemma}
\label{lemma_3} \emph{Let $\mb{X}$ and $\mb{Y}$ be two distinct,
Hermitian and positive matrices of size $n$: $\mb{X}=\mb{X}^H \succ
0$, $\mb{Y}=\mb{Y}^H \succ 0$ and $\mb{X} \neq \mb{Y}$. Then $
\mathrm{Tr} [(\mb{X}-\mb{Y})(\mb{Y}^{-1}-\mb{X}^{-1})] \geq 0$.}
\end{lemma}
\begin{proof}
It is easy to see that $ \mathrm{Tr}
[(\mb{X}-\mb{Y})(\mb{Y}^{-1}-\mb{X}^{-1})]=\mathrm{Tr}[ \mb{Z} +
\mb{Z}^{-1}-2\mb{I}]$, with the Hermitian and positive matrix
$\mb{Z} \triangleq
\mb{X}^{\frac{1}{2}}\mb{Y}^{-1}\mb{X}^{\frac{1}{2}}$ and thus we
further have  $ \mathrm{Tr}
[(\mb{X}-\mb{Y})(\mb{Y}^{-1}-\mb{X}^{-1})]=\ds{ \sum_{i=1}^n
\frac{(\lambda_Z(i)-1)^2}{\lambda_Z(i)}} \geq 0$
where the matrix $\mb{\Lambda}_Z = \mathrm{Diag}(\lambda_Z(1), ...,
\lambda_Z(n))$ corresponds to the spectral decomposition of .
\end{proof}
By applying this lemma to $\mc{T}$ we have that:
\begin{equation}
\begin{array}{lcl}
\mc{T} & =& \mathrm{Tr}
\left\{(\mb{A}-\mb{B})(\mb{B}^{-1}-\mb{A}^{-1}) +
[(\mb{C}+\mb{A})-(\mb{B}+\mb{D})][(\mb{B} +
\mb{D})^{-1}-(\mb{A}+\mb{C})^{-1}]  \right.\\
& & \left.- (\mb{A}-\mb{B})[(\mb{B} +
\mb{D})^{-1}-(\mb{A}+\mb{C})^{-1}] \right\}\\
& \geq & \mathrm{Tr} \left\{(\mb{A}-\mb{B})(\mb{B}^{-1}-\mb{A}^{-1})
 - (\mb{A}-\mb{B})[(\mb{B} + \mb{D})^{-1}-(\mb{A}+\mb{C})^{-1}]
\right\}.
\end{array}
\end{equation}
We know that $\mb{C}\preceq \mb{D}$ then $\mb{C}+\mb{A}\preceq
\mb{D}+\mb{A}$ and also that $(\mb{C}+\mb{A})^{-1}\succeq
(\mb{D}+\mb{A})^{-1}$. Using the fact that $\mb{A}\succeq \mb{B}$
and also Lemma \ref{trace_prod} we have that $
\mathrm{Tr}[(\mb{A}-\mb{B})(\mb{C}+\mb{A})^{-1}]\geq
\mathrm{Tr}[(\mb{A}-\mb{B})(\mb{D}+\mb{A})^{-1}]$ and the trace
becomes lower bounded as $\mc{T} \geq  \mathrm{Tr}
\left\{(\mb{A}-\mb{B})(\mb{B}^{-1}-\mb{A}^{-1})
 - (\mb{A}-\mb{B})[(\mb{B} +
\mb{D})^{-1}-(\mb{A}+\mb{D})^{-1}] \right\}$. Now, we are going to
prove that this lower bound, say $\mc{T}_{LB}$, is positive:
\begin{equation}
\begin{array}{lcl}
\mc{T}_{LB}&= &\mathrm{Tr}
\left\{(\mb{A}-\mb{B})(\mb{B}^{-1}-\mb{A}^{-1})
 - \left[(\mb{A}+\mb{D})-(\mb{B}+\mb{D})\right]\left[(\mb{B} +
\mb{D})^{-1}-(\mb{A}+\mb{D})^{-1}\right]\right\} \\
& = &
\mathrm{Tr}\left\{(\widetilde{\mb{A}}-\widetilde{\mb{B}})(\widetilde{\mb{B}}^{-1}-\widetilde{\mb{A}}^{-1})
 - \left[(\widetilde{\mb{A}}+\mb{I})-(\widetilde{\mb{B}}+\mb{I})\right]\left[(\widetilde{\mb{B}} +
\mb{I})^{-1}-(\widetilde{\mb{A}}+\mb{I})^{-1}\right]\right\}
\end{array}
\end{equation}
where we have made the following change of variables:
$\mb{A}=\mb{D}^{1/2}\widetilde{\mb{A}}\mb{D}^{1/2}$,
$\mb{B}=\mb{D}^{1/2}\widetilde{\mb{B}}\mb{D}^{1/2}$ such that
$\widetilde{\mb{A}}=\mb{D}^{-1/2}\mb{A}\mb{D}^{-1/2}=\widetilde{\mb{A}}^H
\succ 0$ and
$\widetilde{\mb{B}}=\mb{D}^{-1/2}\mb{B}\mb{D}^{-1/2}=\widetilde{\mb{B}}^H
\succ 0$. By applying the Woodbury formula
$(\widetilde{\mb{A}}+\mb{I})^{-1}=\widetilde{\mb{A}}^{-1}-\widetilde{\mb{A}}^{-1}
(\widetilde{\mb{A}}+\mb{I})^{-1}$ and
$(\widetilde{\mb{B}}+\mb{I})^{-1}=\widetilde{\mb{B}}^{-1}
-\widetilde{\mb{B}}^{-1}(\widetilde{\mb{B}}+\mb{I})^{-1}$, the lower
bound $\mc{T}_{LB}^{(1)}$ rewrites as:
\begin{equation}
\begin{array}{lcl}
\mc{T}_{LB}&= &
\mathrm{Tr}\left\{(\widetilde{\mb{A}}-\widetilde{\mb{B}})\left[\widetilde{\mb{B}}^{-1}-\widetilde{\mb{A}}^{-1}
-\widetilde{\mb{B}}^{-1}+\widetilde{\mb{B}}^{-1}
(\widetilde{\mb{B}}+\mb{I})^{-1}+\widetilde{\mb{A}}^{-1}-\widetilde{\mb{A}}^{-1}
(\widetilde{\mb{A}}+\mb{I})^{-1}\right]\right\}
\\
& = & \mathrm{Tr}\left\{
\widetilde{\mb{A}}\widetilde{\mb{B}}^{-1}(\widetilde{\mb{B}}+\mb{I})^{-1}
+ \widetilde{\mb{B}}
\widetilde{\mb{A}}^{-1}(\widetilde{\mb{A}}+\mb{I})^{-1} -
(\widetilde{\mb{A}}+\mb{I})^{-1}-
(\widetilde{\mb{B}}+\mb{I})^{-1}\right\}.
\end{array}
\end{equation}
Let us denote the ordered eigenvalues of the two matrices
$\widetilde{\mb{A}}$ and $\widetilde{\mb{B}}$ as
$\lambda_{\widetilde{A}}(1)\leq \lambda_{\widetilde{A}}(2) \leq
\hdots \leq \lambda_{\widetilde{A}}(n)$ and
$\lambda_{\widetilde{B}}(1)\leq \lambda_{\widetilde{B}}(2) \leq
\hdots \leq \lambda_{\widetilde{B}}(n)$. From
\cite{lasserre-ac-1995} we know that for two matrices $\mb{X}$ and
$\mb{Y}$ of size $n$, $\mathrm{Tr}(\mb{X} \mb{Y}) \geq
\sum_{i=1}^{n} \lambda_X(i) \lambda_Y(n-i+1)$, which implies
directly that $\mathrm{Tr}(\mb{X} \mb{Y}^{-1}) \geq \sum_{i=1}^{n}
\frac{\lambda_X(i)}{\lambda_Y(i)}$, where $\lambda_X(i)$ and
$\lambda_Y(i)$ are the ordered eigenvalues (in the previously
specified order) of the corresponding matrices. Applying this result
we find that
\begin{equation}
\begin{array}{lcllcl}
 \mathrm{Tr}\left[\widetilde{\mb{A}}
 \widetilde{\mb{B}}^{-1}(\mb{I}+\widetilde{\mb{B}})^{-1}
 \right] & \geq & \ds{\sum_{i=1}^{n} \frac{\lambda_{\widetilde{A}}(i)}
  {\lambda_{\widetilde{B}}(i)(1+\lambda_{\widetilde{B}}(i))}}, &
 \mathrm{Tr}\left[\widetilde{\mb{B}}\widetilde{\mb{A}}^{-1}
 (\mb{I}+\widetilde{\mb{A}})^{-1}\right] & \geq &
  \ds{\sum_{i=1}^{n} \frac{\lambda_{\widetilde{B}}(i)}
  {\lambda_{\widetilde{A}}(i)(1+\lambda_{\widetilde{A}}(i))}},
\end{array}
\end{equation}
and finally obtain that:
\begin{equation}
\begin{array}{lcl}
\mc{T}_{LB}&\geq 
&\ds{\sum_{i=1}^{n}
\frac{[\lambda_{\widetilde{A}}(i)-\lambda_{\widetilde{B}}(i)]^2
[1+\lambda_{\widetilde{A}}(i)+\lambda_{\widetilde{B}}(i)]}{\lambda_{\widetilde{A}}(i)
\lambda_{\widetilde{B}}(i)[1+\lambda_{\widetilde{A}}(i)][1+\lambda_{\widetilde{B}}(i)]}}
 \geq 0.
\end{array}
\end{equation}

To conclude the global proof one can easily check that Case (2)
(resp. Case (4)) can be readily proved from the proof of Case (1)
(resp. Case (3)) by interchanging the role of $\mb{A}$ and $\mb{B}$
and $\mb{C}$ and $\mb{D}$.


\bibliography{biblio}



\begin{figure}
  \begin{center}
    \includegraphics[scale=0.60]{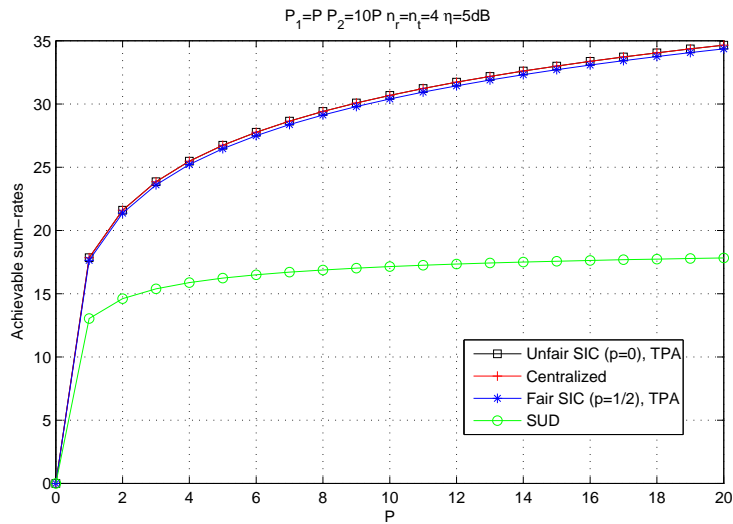}
  \end{center}
  \caption{ \footnotesize Temporal PA game. MAC sum-rate versus
  the transmit power $P$
  for $P_1=P$, $P_2=10P$, $n_r=n_t=4$, $\eta=5$ dB. Comparison between the fair SIC decoding scheme
  ($p=\frac{1}{2}$), the unfair SIC scheme ($p=0$), and SUD decoding scheme.}
  \label{fig2}
\end{figure}

\begin{figure}
  \begin{center}
      \includegraphics[scale=0.60]{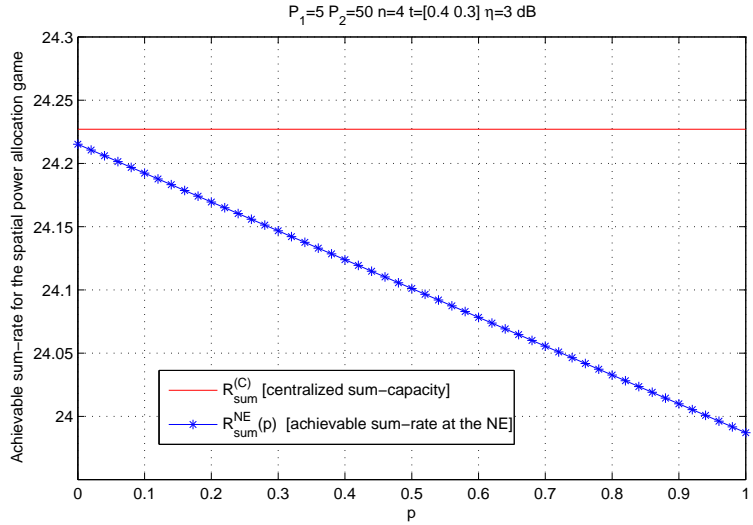}
  \end{center}
  \caption{ \footnotesize Spatial PA game. MAC sum-rate versus $p$
  for $P_1=5$, $P_2=50$, $n_r=n_t=4$, $\eta=3$ dB, $t_1=0.4$, $t_2=0.3$. The achievable network sum-rate of
  fading MIMO MACs is linear w.r.t. $p \in [0,1]$ and is very close to
  the centralized upper bound. The optimal distribution obtained with the Stackelberg game is $p^*=0$.}
  \label{fig3}
\end{figure}

\begin{figure}
  \begin{center}
      \includegraphics[scale=0.60]{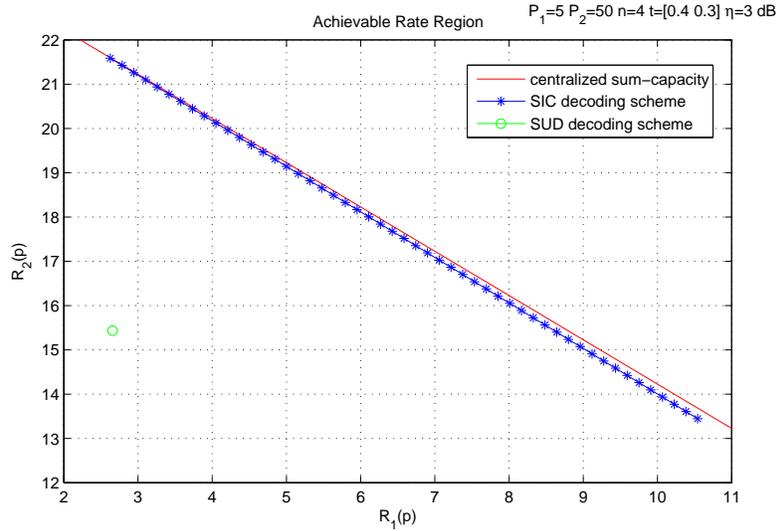}
  \end{center}
  \caption{ \footnotesize Spatial PA game. Achievable rate
  region
   for $P_1=5$, $P_2=50$, $n_r=n_t=4$, $\eta=3$ dB, $t_1=0.4$, $t_2=0.3$. By varying $p$
   allows to move along a segment close to the centralized sum-capacity, similar to
   the SISO MAC channels.}
  \label{fig4}
\end{figure}


\end{document}